\documentclass{emulateapj}
\usepackage{color}
\usepackage{graphicx}
\usepackage{dcolumn}
\usepackage{bm}
\begin{document}

\shorttitle{Shapes and orientations of core-S\'ersic galaxies }
\title{Core shapes and orientations of core-S\'ersic galaxies}

\shortauthors{Dullo \& Graham} 
\author{Bililign T.\ Dullo\altaffilmark{1},  Alister W.\ Graham\altaffilmark{1}}
\affil{\altaffilmark{1} Centre for Astrophysics and Supercomputing, Swinburne University
of Technology, Hawthorn, Victoria 3122, Australia; Bdullo@astro.swin.edu.au}

\begin{abstract}
The inner and outer shapes and orientations of core-S\'ersic galaxies may hold
important clues to their formation and evolution.  We have therefore
measured the central and outer ellipticities and position angles for a sample of 24 
core-S\'ersic  galaxies using
archival {\it HST} images and data. By selecting galaxies with core-S\'ersic break radii $R_{b}$---a measure
of the size of their partially depleted core--- that are $\ga 0\arcsec$.2,  we find that the ellipticities
 and position angles are quite robust against {\it HST} seeing.   For the bulk of the galaxies, there is 
 a good agreement between the  ellipticities and position
angles at the break radii and the average outer ellipticities and
position angles determined over $R_{\rm e}/2 < R <R_{\rm e}$, where $R_{\rm e}$ is
the spheroids' effective half light radius. However there are some interesting differences. We find a median ``inner"
ellipticity at $R_{b}$ of $\epsilon_{\rm med}$= $0.13 \pm
0.01$,  rounder than the median ellipticity of the ``outer" regions
 $\epsilon_{\rm med}$= $0.20\pm 0.01$, which is thought to reflect  the influence of the 
 central supermassive black hole at
small radii. In addition, for the first time we find a trend, albeit weak (2 $\sigma$ significance), such that galaxies with
larger (stellar deficit)-to-(supermassive black hole) mass ratios---thought to be
 a measure of the number of major dry merger events---tend to have rounder inner and outer isophotes, suggesting a  connection
between the galaxy shapes and their merger histories. We show that this finding is not 
simply reflecting the well known result that more luminous galaxies are rounder, but it is no doubt related.

\end{abstract}


\section{Introduction}

In the hierarchical structure formation model, massive early-type
galaxies are believed to be formed via galaxy mergers (e.g., Toomre \&
Toomre 1972; Schweizer 1982; Barnes 1988; Kauffmann, White \&
Guiderdoni 1993). The most favored core-S\'ersic galaxy formation
scenario predicts the creation  of their partially depleted cores by
the action of coalescing supermassive black hole (SMBH) binaries within 
the merged remnant of their gas-poor progenitors.  The
energy which is transferred  from the SMBH binary's orbital decay
causes the slingshot ejection of the inner stars, creating the stellar light
deficits of core-S\'ersic galaxies (e.g., Begelman et al.\ 1980;
Ebisuzaki et al.\ 1991; Milosavljevi\'c \& Merritt 2001; Merritt 2006).
This three-body scattering process preferentially removes stars on
radial (eccentric) orbits, resulting in tangentially biased (circular)
orbits in the core region of core-S\'ersic galaxies, and excess levels
of radial orbits beyond the core (e.g., Quinlan \& Hernquist 1997;
Milosavljevi\'c \& Merritt 2001; Gebhardt et al.\ 2003). Thomas et
al.\ (2014) have recently found these kinematical signatures in six
core-S\'ersic galaxies. It, therefore, seems reasonable that the inner
isophotal shape, orientation (position angle) and radial light distribution of a
core-S\'ersic galaxy may be connected to each other and to the details
of the physical processes that built the galaxy at large. For instance, does the orientation
of the core still reflect the orbital plane of the merging SMBHs, and
does that also reflect the orbital alignment of the pre-merged
galaxies and the final outer/global isophotal shape of the wedded
galaxy pair? Such questions are the main motivation for this work.

``Core-S\'ersic" galaxies have central 
light deficits relative to the
(steeper) inward extrapolation of their spheroids' (i.e., elliptical
galaxies' or bulges') outer S\'ersic profile (Graham et al.\ 2003), while 
``core" galaxies with an inner slope $\gamma < 0.3$ (Lauer et al.\ 1995) 
do not necessarily have deficits relative to the outer S\'ersic profile. The
 Nuker model break radii are typically $2-3$ times larger than the 
core-S\'ersic model break radii (e.g., Trujillo et al.\ 2004); i.e., they are
 somewhat outside of the main depleted core and therefore not an ideal 
 location to measure the core's ellipticity and position angle. However, the 
 smaller radius where the negative, logarithmic slope of the projected light 
 profile equals 0.5 can be derived from the Nuker model parameters (Carollo
  et al.\ 1997; Rest et al.\ 2001) and matches very well with the core-S\'ersic break radius
   (Dullo \& Graham 2012). However, Dullo \& Graham (2012)
 found that $\sim$18\% of ``cores" 
 according to the Nuker model are not depleted cores according to the 
 core-S\'ersic model and as such, application of the core-S\'ersic model is still required
  if one wishes to identify galaxies with partially depleted cores rather than simply flattish inner (non-depleted)
   S\'ersic light profiles. Graham (2004) asked
whether the ellipticity (and position angle) at the break radii of actual partially depleted cores 
in core-S\'ersic galaxies agrees and aligns with the outer spheroids'
ellipticity (and position angle). With a sufficiently large (N=24) sample
 we can now start to probe this question.

Previous works on the
evolution of triaxial galaxies had indicated that a central
supermassive black hole  would destroy the inner box
orbits leaving a rounder or nearly spherical core (e.g., Norman, May,
\& van Albada 1985; Gerhard \& Binney 1985; Merritt \& Quinlan 1998;
Merritt 2000; Holley-Bockelmann et al.\ 2002; Merritt \& Poon 2004), while
the outer parts of core-S\'ersic spheroids are
thought to be built, shaped and oriented by dissipationless, major
galaxy merging and the ensuing strong violent relaxation and phase
mixing (e.g., van Albada 1982; McGlynn 1984; Burkert et al.\ 2008;
Hopkins et al.\ 2009). In addition, the change in the shapes and
orientations of the isophotes at large radii may arise from the effect
of tidal encounters with nearby massive companions (Nieto \& Bender
1989; Hao et al.\ 2006), and the presence of isophotal twisting
in galaxies are also commonly associated with triaxiality, bars, disks or
dust lanes (e.g., Nieto et al.\ 1992). In this paper, we investigate the
radial changes of the shapes and orientations of core-S\'ersic
galaxies, and the implications for their formation and evolution. We 
note that core-S\'ersic galaxies are not confined to elliptical galaxies. 
Lenticular disc galaxies, i.e., ``fast rotators", are also known to contain 
partially-depleted cores relative to their bulge's main S\'ersic profile (Dullo \& Graham 2013; Krajnovi\'c et al.\ 2013).
 While these discs may be  associated with the galaxy's previous merger event 
 (e.g., Naab \& Burkert 2003), they may have been subsequently accreted from cold gas flows, 
 possibly fed along streams of a fixed orientation (Pichon et al.\ 2011).
  In what follows we use the shapes of
  the bulge, or bulge-dominated, region of the galaxies.

In Section~\ref{SecV2} we describe our data selection, and isophotal
profile extraction techniques. The measurements of the inner and outer
ellipticities and orientations of the galaxies are presented in
Sections~\ref{SecV3} and \ref{SecV4},
respectively. Section~\ref{SecV5} discusses the implications of these
results in the context of core-S\'ersic galaxy formation and
evolution. We also compare our study with previous works. Our main conclusions are summarized in Section~\ref{ConV}.

 \section{Galaxy Sample and Profiles}\label{SecV2}

We have used the IRAF task {\sc ellipse} (Jedrzejewski 1987) to
extract ellipticity and position angle profiles from {\it HST} WFPC2
and ACS images of a sample of 24  nearby ($\sim$10-100 Mpc)
core-S\'ersic galaxies with $M_{V} \la -20.7$ mag presented in Dullo \&
Graham (2014). These images were retrieved from the Hubble Legacy
Archive\footnote{http://hla.stsci.edu}.  All the 24 galaxies in our sample have
 partially depleted cores relative to their outer
S\'ersic profile (see Table~\ref{TabV1}). They have come from an
initial sample of 39 galaxies (Lauer et al.\ 2005) with shallow inner
profile slopes, but not necessarily depleted cores (see Dullo \&
Graham 2012, 2013). Excluding seven S\'ersic galaxies (which do not have depleted cores) and one dusty galaxy, 
in Dullo \& Graham (2014) we studied 31 of these core-S\'ersic galaxies.
Here, we excluded seven core-S\'ersic galaxies from the Dullo \&
Graham (2014) sample. Specifically,  NGC 1700, NGC 3640 and
NGC 7785 because they have questionably  small cores ($R_{b}\la 0\arcsec.04$, Dullo
\& Graham 2014),  and as such the ellipticity and position angle measurements at
their break radii cannot be trusted. For NGC 3706, NGC 4073, NGC 4406 and NGC 6876,
Lauer et al.\ (2005) did not publish the inner ($R \la1\arcsec$)
isophote ellipticity and position angle profiles because their {\sc ellipse}
fits were unstable for these inner isophotes. 

To check on the influence of
the {\it HST}'s Point Spread Function (PSF), we have additionally used the
inner ($R \la 3\arcsec$) Lauer et al.\ (2005) PSF-deconvolved
ellipticities and position
angles\footnote{http://www.noao.edu/noao/staff/lauer/wfpc2\_profs/}. Details
of our data reduction and isophotal parameter extraction procedures
are discussed in Dullo \& Graham (2013). 

For comparison, we (Bonfini et al.\ 2014, in prep.) also derived
ellipticity and position angle profiles by fitting PSF-convolved 2D
core-S\'ersic models to the galaxy images using GALFIT (Peng et
al.\ 2002; Bonfini 2014).  In general, our ellipticities and position
angles determined from the 1D {\sc ellipse} fits agree reasonably well with i) the
results based on GALFIT (see also Floyd et al.\ 2008 for similar
conclusions) and ii) those from Lauer et al.\ (2005) for $R \ga
0\arcsec.2$ and out to their outer most radii of $\approx 10 -15\arcsec$. Here we 
use profiles out to $\approx 100\arcsec$, typically extending beyond the galaxies
 half light radii.  In this work, we use the ellipticities and
position angles from the {\sc ellipse} fits because GALFIT fits
PSF-convolved elliptical core-S\'ersic models with fixed (i.e., singular)
ellipticities and position angles to the galaxy images. However, the
shapes and orientations of galaxies can of course change with
radius  (e.g., di Tullio 1979; Kent 1984;
Zaritsky \& Lo 1986, Bender et al.\ 1988; Capaccioli et
al.\ 1988). Table~\ref{TabV1} presents the morphological classification, core-S\'ersic 
break radii, spheroids' effective half light radii, ellipiticities and position angles for our sample
galaxies\footnote{The average values were derived using a logarithmic sampling of the radius.}.
 
 \begin{figure}
\includegraphics[angle=270,scale=0.86]{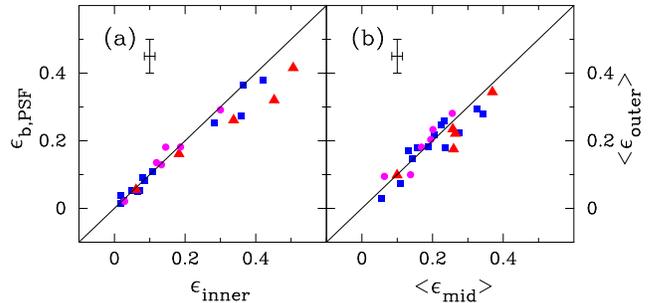}
\caption{Panel (a): PSF-affected ellipticity measured at the core-S\'ersic break
  radius ($\epsilon_{b,\rm PSF}$) plotted against the Lauer et
  al.\ (2005) PSF-deconvolved ellipticity at the core-S\'ersic break
  radius ($\epsilon_{\rm inner}$). Panel (b): average ellipticity
  determined at $R > R_{\rm e}$ ($<$$\epsilon_{\rm outer}$$>$) versus
  average ellipticity measured between $R_{\rm e}/2$ and $R_{\rm e}$
  ($<$$\epsilon_{\rm mid}$$>$). Circles and squares
  represent core-S\'ersic elliptical galaxies with and without
  additional nuclear light components, respectively. Triangles
  designate core-S\'ersic lenticular galaxies.}
\label{FigV2} 
 \end{figure}

\section{Ellipticities}\label{SecV3} 
\subsection{Assessment of the ellipticity measurements}\label{secV3.1}

In Fig.~\ref{FigV2}a, we test the effect of the {\it HST} WFPC2/ACS PSF
by comparing our ellipticities determined from the {\sc ellipse} fits with
the Lauer et al.\ (2005) PSF-deconvolved ellipticities, both measured (by us)
at the core-S\'ersic break radius. The {\it HST}'s WFPC2/ACS seeing
has a full width at half-maximum, FWHM $\approx 0\arcsec.1$. In
general, PSF smearing biases inner galaxy isophotes towards rounder
ellipticties (e.g., Rest et al.\ 2001; Rahman \& Shandarin 2003;
Holden et al.\ 2009).  Simulations by Holden et al.\ (2009), for
example, showed that {\it HST}'s PSF causes the apparent ellipticities
of high redshift ($z >0.3$) early-type galaxies to be
underestimated. Nonetheless, the ``break
ellipticities\footnote{Throughout this paper, the terms ``break
  ellipticity'' and ``break position angle'' refer to the measurements
  of the ellipticity and position angle at the core-S\'ersic break radius
  (Table~\ref{TabV1}).}''  from our {\sc ellipse} fit ($\epsilon_{b,
  \rm PSF}$) and those from Lauer et al.\ (2005), $\epsilon_{\rm
  inner}$, agree within the errors,
indicating minimum PSF contamination at our galaxy sample's  break radii, which are all $>0\arcsec.2$.

As can be seen in Fig.~\ref{FigV2}a, there is a systematic tendency for the 
ellipticity of the most elliptical cores to be slightly under-estimated from the {\sc ellipse} fits. 
This is likely due to seeing.  Throughout this paper we use the Lauer et al.\ (2005) PSF-deconvolved
ellipticity profile sampled at our core-S\'ersic break radius  ($\epsilon_{\rm inner}$) 
and the average spheroid ellipticity measured by us using our {\sc ellipse} data over the
interval $R_{\rm e}/2$ to $R_{\rm e}$ (denoted $<$$\epsilon_{\rm mid}$$>$) as the
spheroid's representative inner and mid-range ellipticities, respectively
(Table~\ref{TabV1}).  Note that our average  ellipticities and
position angles (Table~\ref{TabV1}) are not luminosity weighted.

Six of our 24 galaxies (NGC 0741, NGC 4278, NGC 4365, N4472, NGC 4552
and NGC 5419) have additional
nuclear light components (Dullo \& Graham 2012). For four of these 
nucleated galaxies (i.e., except for NGC 4365 and NGC 5419), both $\epsilon_{\rm
  inner}$ and $<$$\epsilon_{\rm mid}$$>$ are lower than $\la 0.20$ and
their central light excess is due to a point-source AGN (Dullo \&
Graham 2012, their Section 8). For NGC 5419,  Lena et al.\ (2014) noted that it
has a double nucleus (see also Lauer et al.\ 2005). They found that the galaxy 
photocenter is displaced by $\sim  $7.5 pc and 62 pc  (i.e., $\la 0.15 R_{b,\rm N5419}$)
  from these nuclei. Thus, the double nucleus is unlikely to affect our measurement of
  the ellipticity at this galaxy's core-S\'ersic break radius $R_{b,\rm N5419}$=416
   pc (Dullo \& Graham 2014). The elliptical galaxy NGC 4365, with
$\epsilon_{\rm inner}$ $\sim 0.30$ and $<$$\epsilon_{\rm mid}$$>$
$\sim 0.26$, has an elongated nuclear stellar cluster (e.g., Carollo
et al.\ 1997; Lauer et al.\ 2005; C\^ot\'e et al.\ 2006; Dullo \&
Graham 2012). Carollo et al.\ (1997) noted that this nucleus  extends 
from 1$\arcsec$ to $3\arcsec$, compared to this galaxy's core-S\'ersic break radius
 $R_{b,\rm N4365}$=1$\arcsec.21$ (Dullo \& Graham 2014). Therefore,  
 our measurements of $\epsilon_{\rm inner}$ and  $<$$\epsilon_{\rm mid}$$>$ 
 for NGC 4365 are likely to be influenced by its nucleus.

In order to assess the stability of the outer ellipticities, in
Fig.~\ref{FigV2}b, we compare $<$$\epsilon_{\rm mid}$$>$ with the
average spheroid ellipticity between $R_{\rm e}$ and the outermost
 ($R \approx 100\arcsec$) data point ($<$$\epsilon_{\rm
  outer}$$>$). Overall, Fig.~\ref{FigV2}b shows that the spheroid's
ellipticity profile remains stable over $R_{\rm e}/2 \la R \la
100\arcsec$. Due, however, to potential tidal effects at large radii,
or not fully relaxed outer regions, and the increasing dominance
 of disk light in the lenticular galaxies,  we have elected to use
  $<$$\epsilon_{\rm mid}$$>$ to represent the spheroid outside of its core region.

\begin{figure}
\includegraphics[angle=270,scale=0.86]{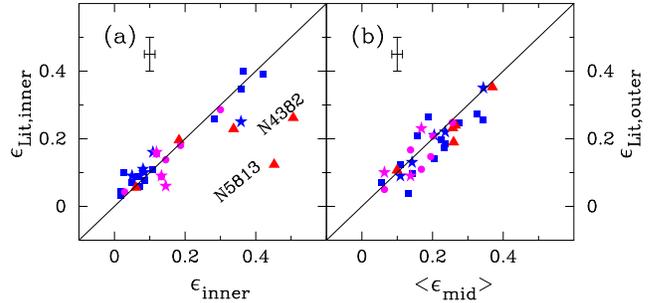}
\caption{Panel (a): our break radius ellipticity $\epsilon_{\rm inner}$ plotted
  against various inner ellipticities from the literature, $\epsilon_{\rm
    lit,\rm inner}$: the Ryden et al.\ (2001, their Table~1) inner
  ($R=R_{\rm e}/16$) ellipticity for 10 core-S\'ersic galaxies
  (filled stars), and inner luminosity-weighted, average ellipticity
  taken from Lauer et al.\ (2005, their Table 5). Panel (b): our
  average ellipticity  between $R_{\rm e}/2$ and $R_{\rm e}$
  ($<$$\epsilon_{\rm mid}$$>$) plotted against  outer
  ellipticities from the literature, $\epsilon_{\rm lit,\rm outer}$: the
  Ryden et al.\ (2001, their Table 1) outer ($R=R_{\rm e}$, Ryden et al.\ 2001) ellipticity
  (filled stars), and outer  ellipticities from Lauer et al.\ (2005, their Table 5). Symbols 
  have the same meaning as in Fig.~\ref{FigV2}. }
\label{FigV2I} 
 \end{figure}

An additional interesting test of our adopted ellipticities is the comparison
to published inner and outer ellipticites by Ryden et al.\ (2001, their
Table 1) and Lauer et al.\ (2005, their Table 5).  Figs.~\ref{FigV2I}a
and \ref{FigV2I}b show that our ellipticity measurements generally
agree well with the differently measured ellipticities from those two works. 
 However,  for NGC 4382 and
NGC 5813, the inner ellipticity at the core-S\'ersic break radius  disagrees with the Lauer et
al.\ (2005) average luminosity weighted ellipticities inside the Nuker
break radius.  These two galaxies have
PSF-deconvolved ellipticity profiles which steadily decrease from
$\sim 0.5$ at $R_{b}$ to $\sim 0.1$ at the galaxies' innermost regions
($R~ \sim0\arcsec.02$). While this might be due to the PSF, Rest et
al.\ (2001) and Lauer et al.\ (1998; 2005) argued that the
deconvolution process can robustly remove the effect of the PSF. 
These gradients are therefore thought to be real.

\begin{table*}
\begin {minipage}{175mm}
\caption{Core-S\'ersic galaxy data.}
\label{TabV1}
\begin{tabular}{@{}llccccccccccc@{}}
\hline
\hline
Galaxy&Type&$R_{b}$&$R_{\rm e}$&$\epsilon_{b,\rm PSF}$&$\epsilon_{\rm inner}$&$<$$\epsilon_{\rm mid}$$>$&$<$$\epsilon_{\rm outer}$$>$&P.A.$_{b,\rm PSF}$&P.A.$_{\rm inner}$&$<$P.A.$_{\rm mid}$$>$&$<$P.A.$_{\rm outer}$$>$&\\
&&(arcsec)&(arcsec)&&&&&(deg)&(deg)&(deg)&(deg)&\\
(1)&(2)&(3)&(4)&(5)&(6)&(7)&(8)&(9)&(10)&(11)&(12)\\
\multicolumn{1}{c}{} \\               
\hline                            
NGC 0507 &S0     &0.33   &5.3&0.261& 0.337    & 0.260   & 0.176&10.8& 20.3  &10.6& 14.8\\
NGC 0584   &E$^{d}$ &0.21   &112.5  &0.272&0.358&0.344    &  0.279& 60.0 &59.2& 68.7  & 58.6       \\
NGC 0741  &E &0.76&53.0   &0.021&0.029  &  0.195  &0.204 & 52.2&125.9   &   89.9& 88.7           \\
NGC 1016 &E   &0.48    &41.7&0.025&  0.025&0.056    &0.030& 80.0& 40.1   &  34.7 & 9.6   \\
NGC 1399&E    &2.30 &36.6   &0.082&0.086&0.109   & 0.075& 112.4 &112.1& 110.0 & 95.7                      \\
NGC 2300&S0  &0.53&7.7    &0.162 & 0.183 & 0.257 & 0.235  & 78.6 &78.6 & 74.5 &73.9\\
NGC 3379  &E &1.21&50.2  &0.092&0.081&0.142    &0.147&  61.1&70.8&  66.5 & 64.5    \\
NGC 3608 &E  &0.23 & 68.7 &0.110& 0.109&0.236 & 0.179&72.9&65.0    &   82.1  & 92.3  \\
NGC 3842   &E      &0.72&102.4    &0.014&0.017 & 0.232 & 0.258& 142.8 &109.0 &  163.1&140.0  \\
NGC 4278 &E    &0.83&20.2    &0.136& 0.119    &0.138  &0.100&18.8& 8.7  &  16.8  &   26.6\\
NGC 4291 &E    &0.30&13.6   & 0.254& 0.282   & 0.276   &0.225 &106.0   &  104.6&104.8 &102.9    \\
NGC 4365 &E   &1.21&47.3   &0.292 & 0.300    &  0.256  &0.281&44.5&45.1 &42.4 &40.7\\
NGC 4382&S0    &0.27  & 11.1 &0.416&0.506     &  0.264   &0.221& 43.2 & 43.2 &30.7 & 27.9          \\
NGC 4472 &E$^{d}$&1.21  &  48.8&0.129& 0.133 & 0.169& 0.181 & 161.0&160.3  & 159.1   &157.7    \\
NGC 4552 &E$^{d}$  &0.38 &29.7  &0.181 &0.145  & 0.064& 0.095 &136.3&132.3&   124.7 &131.4      \\
NGC 4589&E   &0.20   &70.8 &0.366 & 0.365  &0.224& 0.248 & 112.8& 112.9 & 83.7&74.5  \\
NGC 4649&E     &2.51&   62.8 &0.053&0.050   &  0.204 &0.218 & 93.6&  91.2 & 101.8    &  104.1         \\
NGC 5061&E   &0.22 &68.4  &0.053 & 0.072  &0.132&0.171& 179.8&175.0  & 110.6 & 113.6    \\
NGC 5419&E$^{d}$&1.43 &55.0   &0.182& 0.187& 0.201&   0.233&92.9&93.6 & 77.9  & 72.1  \\
NGC 5557&E    &0.23&  30.2&0.050  &0.067   & 0.156& 0.180&   87.9& 79.4 &  95.7 &88.8       \\
NGC 5813 &S0    &0.35 &7.1   &0.321   & 0.452  & 0.100&0.099& 134.9 &140.4& 148.1& 139.1\\
NGC 5982&E     &0.25&  26.8  &0.038& 0.019   &0.327 & 0.295 & 124.2&92.8&   108.7 &107.2     \\
NGC 6849&SB0    &0.18&7.8 &0.056& 0.061 & 0.367&0.344& 8.4&163.5& 24.0 &23.2 \\
NGC 7619&E     &0.49 &72.2  &0.381& 0.420   &  0.189 &  0.181&35.3&35.7 &   40.4 &39.4\\
\hline
\end{tabular} 

Notes.---Col.\ (1) Galaxy name. Col.\ (2) Morphological classification
taken from Dullo \& Graham (2014). The
superscript $d$ shows galaxies initially classified as ellipticals but which have recently been re-classified as disc galaxies in the literature by Laurikainen et al.\ (2010). Col.\ (3) Core-S\'ersic break
radius (Dullo \& Graham 2014, their Table 2). Col.\ (4)  The spheroid effective half light radius  (Dullo \& Graham 2014, their Table 2). Col.\ (5) PSF-affected ellipticity at the core-S\'ersic break radius determined using
our {\sc ellipse} fit data. Col.\ (6) The Lauer et
al. (2005) PSF-deconvolved ellipticity at the core-S\'ersic break
radius $R_{b}$. Cols.\ (7) and (8) Average  ellipticities measured using
our {\sc ellipse} fit data over the intervals $R_{\rm e}/2 < R< R_{\rm e}$  and $R > R_{\rm e}$, respectively. Col.\ (9) PSF-affected position angle at the core-S\'ersic break radius.  Col.\ (10) The Lauer et al.\ (2005) PSF-deconvolved position
angle at $R_{b}$. Cols.\ (11) and (12) Average  position angles over
$R_{\rm e}/2 < R< R_{\rm e}$ and $R > R_{\rm e}$, respectively. Position angles of the isophotes are measured
from north to east.
\end {minipage}
\end{table*}

\subsection{Inner and outer core-S\'ersic galaxy ellipticities}\label{secV3.2}

\begin{figure}
\includegraphics[angle=270,scale=0.49]{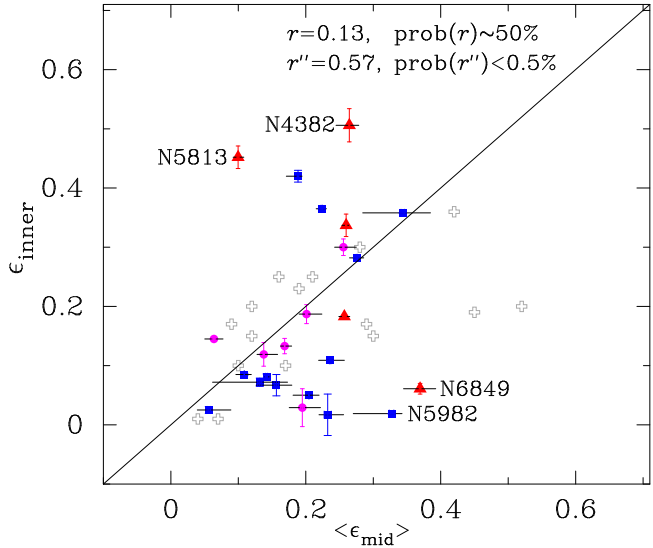}
\caption{Top panel: ellipticity  at the core-S\'ersic break radius ($\epsilon_{\rm inner}$)  versus the 
  average ellipticity between $R_{\rm e}/2 $ and $R_{\rm e}$  ($<\epsilon_{\rm mid}>$). The
  solid one-to-one line is shown for comparison. The values $r$ and prob($r$) are
  the Pearson correlation coefficient and the associated probability,
  respectively, while $r''$ and prob($r''$) are the Pearson
  coefficient and its probability after excluding three outliers (NGC
  5813, NGC 5982 and NGC 6849). Solid horizontal lines
  connect the highest and lowest galaxy ellipticities between $R_{\rm e}$
  and $R_{\rm e}/2$.  Open crosses represent ``core" galaxies taken from 
  Ryden et al.\ (2001, their Table 1), the remaining symbols are the same as in Fig.~\ref{FigV2}. }
\label{FigV1} 
 \end{figure}

Fig.~\ref{FigV1} plots the Lauer et al.\ (2005) PSF-deconvolved
ellipticity at our core-S\'ersic break radius ($\epsilon_{\rm inner}$)
against the average ellipticity between $R_{e}/2$ and $R_{e}$
($<$$\epsilon_{\rm mid}$$>$) for 24 core-S\'ersic spheroids. The solid horizontal lines connect the lowest
and the highest ellipticity values between $R_{\rm e}/2$ and $R_{\rm e}$ for
the individual galaxies.  At first glance there is no obvious correlation between the
core ellipticity ($\epsilon_{\rm inner}$) and spheroid ellipticity
($<$$\epsilon_{\rm mid}$$>$), the Pearson correlation coefficient for
this distribution is $r \approx 0.13$ with the corresponding
probability of happening by chance $P(r) \approx 50\%$. However,
 removing three outliers (NGC 5813, NGC 5982 and NGC 6849) reveals a general trend of broadly similar ellipticities. 

The $\epsilon_{\rm
  inner}-$$<$$\epsilon_{\rm mid}$$>$ correlation becomes even more
pronounced when we include the Ryden et al.\ (2001, their Table 1)
inner $\epsilon(R=R_{\rm e}/16)$ and outer $\epsilon(R=R_{\rm e})$
ellipticities for 16 additional bright galaxies which were classified 
as ``core" galaxies using the Nuker model (Fig.~\ref{FigV1}, open
crosses). However, the caveats are that i) these ``core" galaxies may
not be equivalent to core-S\'ersic galaxies (Dullo \& Graham 2012) and
ii) the Ryden et al.\ innermost fiducial radius $R_{\rm e}/16$ is about a
factor of three larger than the core-S\'ersic break radius $R_{b}
\approx 0.02 R_{e}$ (C\^ot\'e et al.\ 2007). Nonetheless, comparing
the inner and outer isophote ellipticities of early-type galaxies,
Ryden et al.\ (2001, their Fig.~1) and Lauer et al.\ (2005, their
Fig.~5) highlighted the absence of a systematic variation of the
spheroid ellipticity with radius, a result largely confirmed here. 

\begin{figure}
\includegraphics[angle=270,scale=0.49]{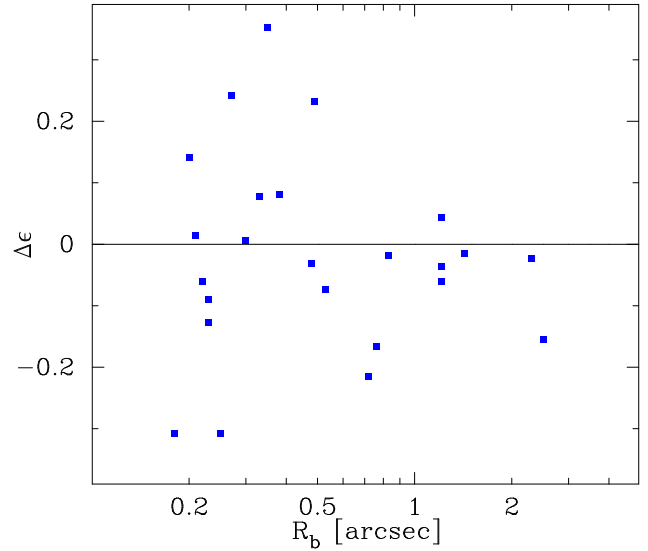}
\caption{Residual of the relation between $\epsilon_{\rm inner}$ and $<$$\epsilon_{\rm mid}$$>$ 
relative to the one-to-one line ($\Delta\epsilon$=$\epsilon_{\rm inner}$$-$$<$$\epsilon_{\rm mid}$$>$, 
Fig.~\ref{FigV1}) 
against core-S\'ersic break radius $R_{b}$. This reveals that the negative $\Delta\epsilon$ of galaxies 
are not because of the influence of seeing on smaller
cores.}
\label{FigV3} 
 \end{figure}

In Fig.~\ref{FigV3}, as a further assessment of the influence of the  PSF, we
explore  $\Delta \epsilon (=\epsilon_{\rm inner}
- $$<$$\epsilon_{\rm mid}$$>$) as a
function of the break radii and find no trend. That is, there is not a trend for galaxies with 
small cores to have rounder cores (a potential artifact of seeing).

\begin{figure}
\includegraphics[angle=270,scale=0.58]{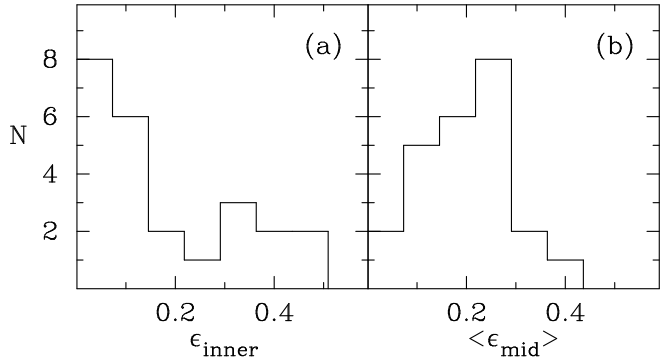}
\caption{Inner and middle (radii) ellipticity distributions for our
  core-S\'ersic galaxies (left and right, respectively). }
\label{FigV1b} 
 \end{figure}

Our ellipticities, $\epsilon_{\rm inner}$
 and $<$$\epsilon_{\rm mid}$$>$,  
 are  broadly consistent with the results from previous
works which showed that  massive, luminous early-type galaxies,
most of which are presumably core-S\'ersic galaxies, tend to be
rounder ($\epsilon \la 0.25$) than their less luminous counterparts,
i.e., the S\'ersic galaxies, $\epsilon \ga 0.25$ (e.g., Davies et
al.\ 1983; Jaffe et al.\ 1994; Ferrarese et al.\ 1994; Faber et
al.\ 1997; Ryden et al.\ 2001; Alam \& Ryden 2002; Vincent \& Ryden
2005; Lauer et al.\ 2005; Ferrarese et al.\ 2006; Emsellem et
al.\ 2007, 2011; Holden et al.\ 2009; Cappellari et al.\ 2011). Overall, our core-S\'ersic galaxies have inner and outer ellipticity
distributions with median ellipticities of
$\epsilon_{\rm med}$ = $0.13 \pm 0.01$, and $\epsilon_{\rm med}$= $0.20\pm 0.01$, respectively
(Fig.~\ref{FigV1b}).  These ellipticities remain the same after excluding the three previously mentioned outliers.  This tendency for core-S\'ersic
galaxies to have $\epsilon_{\rm inner}$  rounder
than $<$$\epsilon_{\rm mid}$$>$ (Fig.~\ref{FigV1b}) may reflect the
influence of the central supermassive black hole at these galaxies' core
regions. However, this does not necessarily imply a trend between  $\epsilon_{\rm inner}$$-$$<$$\epsilon_{\rm mid}$$>$ ( or $\epsilon_{\rm inner}$/$<$$\epsilon_{\rm mid}$$>$) and the supermassive black hole  mass $M_{\rm BH}$ (or mass deficit-to-black hole mass ratio, $M_{\rm def}/M_{\rm BH}$, Section~\ref{Sec3.3}). The quoted errors associated with the median values of $\epsilon_{\rm inner}$ and  $<$$\epsilon_{\rm mid}$$>$ above are the uncertainties on the median, and not the $1\sigma$ scatter. These errors agree with those
 quoted by Holden et al.\ (2009, their Section 3.3). 

\subsection{$M_{\rm def}/M_{\rm BH}$}\label{Sec3.3}
\begin{figure}
\includegraphics[angle=270,scale=0.5]{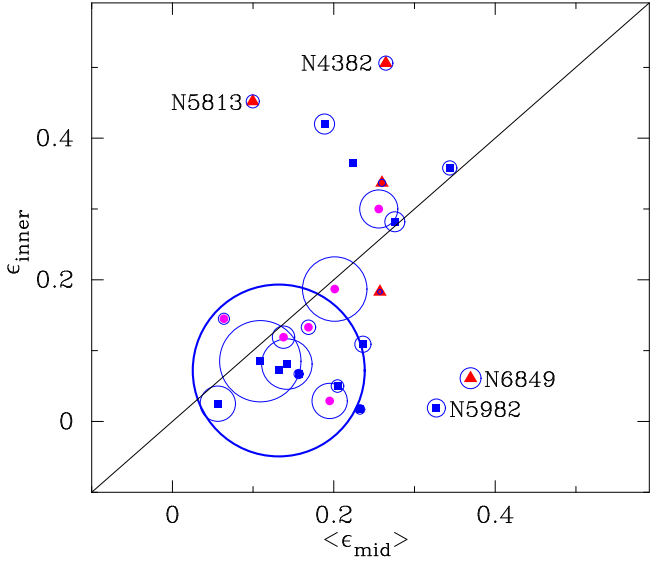}
\caption{Similar to Fig.~\ref{FigV1}, circles designate the (stellar
  mass deficit)-to-(black hole mass) ratios. The sizes of these
  circles increase when the (mass deficit)-to-(black hole mass) ratio
  increases.  Symbols are as in Fig.~\ref{FigV2}. }
\label{FigV1a} 
 \end{figure}

In Fig.~\ref{FigV1a}, open circles represent the ratios of stellar
mass deficit to black hole mass, $M_{\rm def}/M_{\rm BH}~(\approx$$0.5$
$N$, where $N$ is the number of ``dry" major mergers, Merritt 2006) taken from Dullo \& Graham (2014, their Table 4). The
sizes of these circles increase with increasing $M_{\rm def}/M_{\rm
  BH}$ ratio. The ellipticity distribution seen there suggests that
galaxies which have experienced more mergers are rounder (i.e.,
$\epsilon_{\rm inner}$ and $<$$\epsilon_{\rm mid}$$>$ $\la 0.2$). We use the  two-sample  Kolmogorov-Smirnov (KS)  test and the Anderson-Darling (AD)  test (Scholz \& Stephens 1987) to quantify this difference in the ellipticity distribution between the (seven) galaxies with large $M_{\rm def}/M_{\rm BH}$ ratios ($\ge 2.5$) and the (17) galaxies with   $M_{\rm def}/M_{\rm BH} < 2.5$. The results from these two tests agree very well. There is a $\sim$25\% probability that the inner ellipticity ($\epsilon_{\rm inner}$) distributions of galaxies with $M_{\rm def}/M_{\rm BH} \ge 2.5$  and  $M_{\rm def}/M_{\rm BH} < 2.5$ are drawn from the same distribution, while for the mid-radii ellipticity ($<$$\epsilon_{\rm mid}$$>$), this probability decreases to $\sim$10\%.  Further, excluding the nucleated galaxy NGC 4365 (see Section \ref{secV3.1})---which is the only galaxy with both high ellipticities ($\epsilon_{\rm inner}$ and $<$$\epsilon_{\rm mid}$$>$ $> 0.20$) and  $M_{\rm def}/M_{\rm BH} \ge 2.5$---the probabilities corresponding to the inner and mid radii ellipticities drop to $\sim$9\% and $\sim$2.5\%,  respectively. Although the two-dimensional KS test (Peacock 1983; Fasano \& Franceschini 1987) is recommended for sample sizes $\ge 10$, we perform  this test (including NGC 4365) and find a 3.2\% probability that  the ellipticities ($\epsilon_{\rm inner}$, $<$$\epsilon_{\rm mid}$$>$) of galaxies with $M_{\rm def}/M_{\rm BH} \ge 2.5$  and  $M_{\rm def}/M_{\rm BH} < 2.5$ are sampled from the same distribution. Excluding NGC 4365, this drops to 0.3\%.

\begin{figure}
\includegraphics[angle=270,scale=0.5]{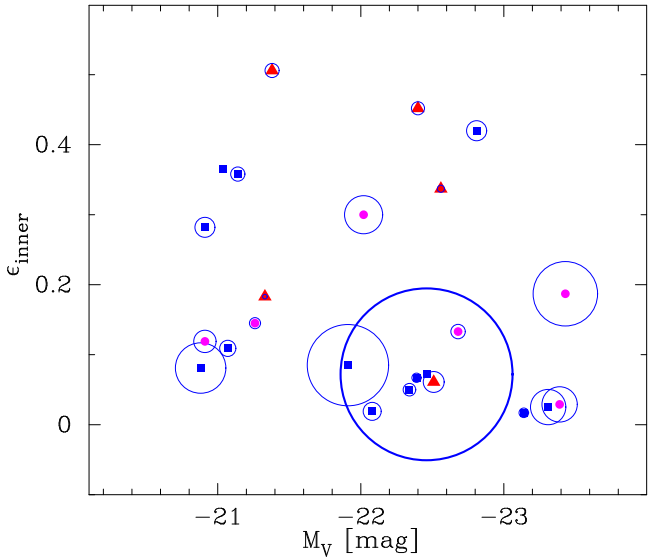}
\caption{Similar to Fig.~\ref{FigV1a}, but now comparing the inner ellipticity ($\epsilon_{\rm inner}$) 
and the spheriod absolute magnitude ($M_{\rm V}$; Dullo \& Graham 2014, their Table 1).}
\label{FigV1aa} 
 \end{figure}

 We have additionally
checked and found that this trend between the galaxy isophotal shape and the $M_{\rm def}/M_{\rm
  BH}$ ratio is not simply because more luminous
galaxies are rounder.  For our core-S\'ersic galaxies, there is no obvious trend between the absolute
magnitude (Dullo \& Graham 2014, their Table 1) and  the ellipticity (Fig.~\ref{FigV1aa}, see also Lauer
et al.\  2005, their Fig.\ 6). The Pearson correlation coefficient for the $\epsilon_{\rm inner}$$-$$M_{\rm V}$ relation (Fig.~\ref{FigV1aa}) is $r \approx$ 0.24 with the corresponding probability $P(r) \approx$ 24\%.

\section{Position angles}\label{SecV4}

\begin{figure}
 \includegraphics[angle=270,scale=0.87]{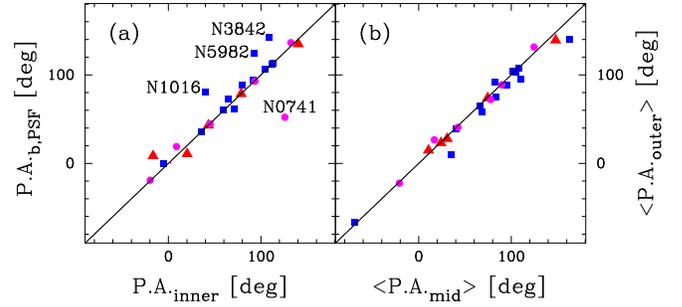}
\caption{Panel (a): comparison of our PSF-affected position angle at
  the core-S\'ersic break radius ($\rm P.A._{b,\rm PSF}$) and the position angle at
  the core-S\'ersic break radius ($\rm P.A._{\rm inner}$) from the PSF-deconvolved 
  profile of Lauer et al.\ (2005).  Panel (b):
  average position angles measured at large radii $R > R_{\rm e}$
  ($<$P.A.$_{\rm outer}$$>$) versus average ellipticity measured between
  $R_{\rm e}$/2 and $R_{\rm e}$ ($<$P.A.$_{\rm mid}$$ >$). Symbols are as in Fig.~\ref{FigV2}. }
\label{FigV4} 
 \end{figure}

\begin{figure}
\includegraphics[angle=270,scale=1.235]{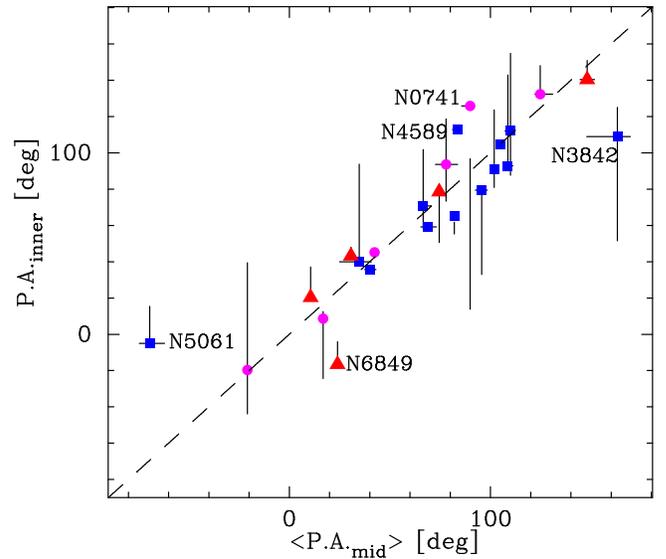}
\caption{Position angle measured at our core-S\'ersic break radius
  ($\rm P.A._{\rm inner}$) versus the average position angle determined
  between $R_{\rm e}/2$ and $R_{\rm e}$ ($<$$\rm P.A._{\rm mid}$$>$). The vertical
  (horizontal) solid line connects two P.A.s which yield the highest
  isophotal twist inside the core (in the range $R_{\rm e}/2 < R <
  R_{\rm e}$). Symbols are as in Fig.~\ref{FigV2}.  }
\label{FigV5} 
 \end{figure}

In this section we evaluate the robustness of our position angle
(P.A.)  measurements, and compare the inner and outer 
orientations (Figs.~\ref{FigV4} and \ref{FigV5}).  In order check on the effect
of the PSF, in Fig.~\ref{FigV4}a we compare the PSF-affected position
angles from our {\sc ellipse} fits with the Lauer et al.\ (2005)
PSF-deconvolved position angle both determined at the core-S\'ersic break radii. The position angles from these two works
agree well except for four galaxies (NGC 0741, NGC 1016, NGC 3842 and
NGC 5982), two of which are discussed below.  
The strong
correlation between the average P.A.s which are determined over the
radial intervals $R_{\rm e}/2< R< R_{\rm e}$ and $R_{\rm e}< R \la100\arcsec$
(Fig~\ref{FigV4}b) implies that the orientations of the galaxies remain, on
average, unchanged between $R_{\rm e}/2$ and $R\approx100\arcsec$.

Using the PSF-deconvolved  P.A.\ profile (Lauer et al.\ 2005), we find that about half of the galaxies in our sample show
isophotal twist higher than $10^{\circ}$ inside their cores, i.e., $R
< R_{b}$, (Fig~\ref{FigV5}, vertical solid lines).
At mid radii ($R_{\rm e}/2<R<R_{\rm e}$) the galaxies
display modest ($\sim 5^{\circ}$) to no twist (Fig~\ref{FigV5}, horizontal solid lines). Lauer et al.\ (2005, their
Fig.\ 9) carried out detailed analyses of isophotal twists for
their ``core" and ``power-law" galaxies\footnote{We refer to ``power-law" 
galaxies  as ``S\'ersic" galaxies because their 
spheroids' light profiles are well described by the curved S\'ersic model rather 
 than by a straight power-law (e.g., Trujillo et al.\ 2004). }  and also concluded that the
amplitude of twists are larger at small radii. This result also holds
for our ``clean" sample of galaxies with partially depleted cores relative to their outer S\'ersic profile,
having also removed galaxies with additional nuclear stellar components.

Also evident in Fig.~\ref{FigV5}, is that despite the twists, the galaxies' orientations at their break
radii are fairly aligned with the spheroid's outer orientation except
for five galaxies (NGC 0741, NGC NGC 3842, NGC 4589, NGC 5061 and NGC
6849) with inner position angles which are twisted by $\ga 27^{\circ}$
from the outer position angle.

NGC 0741, NGC 3842 and NGC 6849 show strong isophotal twists at their
break radii coincident with local minima in their ellipticities, but
for NGC 5061 the strong $30^{\circ}$ twist and the ellipticity minimum
which occurs between $3\arcsec$ and $10\arcsec$ may be
a result of tidal effects (Tal et al.\ 2009). While the correlation
between high twist amplitudes and low isophote ellipticities has
already been known for some time (e.g., Galleta 1980; Nieto et
al.\ 1992; Rahman \& Shandarin 2004), these four galaxies have
interesting features discussed in the literature. NGC 0741 is a
nucleated galaxy (e.g., Dullo \& Graham 2012), NGC 3842 is a bright
cluster galaxy that has an ultramassive black hole ($M_{\rm BH}= 9.7
\times 10^{9} M_{\sun}$) with a 1$\arcsec$.2 radius of influence
(McConnell et al.\ 2011) whereas the S0 galaxy NGC 6849 has an inner
bar which manifests as a peak in the ellipiticity profile (see Dullo \& Graham 2013). The
elliptical galaxy NGC 5061 has the largest S\'ersic index and the
biggest $M_{\rm def}/M_{\rm BH}$ ratio from the Dullo \& Graham
(2014) sample, also, Tal et al.\ (2009) identified a prominent tidal
tail associated with this galaxy.  The remaining outlier in the $\rm
P.A._{\rm inner}-$ $<$$\rm P.A._{\rm mid}$$>$ diagram, NGC 4589 with a dust
 lane in the inner regions, $R\la 10"$,  (e.g., M\"ollenhoff \& Bender 1989; Tomita et al.\ 2000), has
isophotes which smoothly rotate from P.A. $\sim$$112^{\circ}$ at
$R_{b}\sim0.2\arcsec$ to P.A. $\sim85^{\circ}$ at $R_{e}\sim71\arcsec$. 
 This dust lane is visible in the optical {\it HST}/WFPC2 F555W ($V$-band) image.

\section{ Core-S\'ersic galaxy formation, shape and orientation}\label{SecV5} 

Understanding the radial stellar light distributions, and the radial
variation of the shapes and orientations of core-S\'ersic galaxies
(Sections~\ref{SecV3} and \ref{SecV4}) can potentially provide
valuable clues to the details of the mechanisms which drive their
formation and evolution. On the theoretical side, it has been shown that simulated
``dry merger" remnants of spheroids (or bulge-dominated systems) mimic
the slowly rotating massive galaxies with  low
ellipticities (e.g., Khochfar \& Burkert 2003; Boylan-Kolchin et
al. 2006; Naab et al.\ 2006; Naab et al.\ 2007; Burkert et al.\ 2008;
Hopkins et al.\ 2009).  The final shape of a merger remnant depends on
the mass ratio and gas fraction of the progenitors (e.g., Negroponte \& White 1983;
 Barnes 1988; Hernquist 1992;  Naab \&
Burkert 2003; Bournaud, Jog \& Combes 2005; Jesseit, Naab \& Burkert
2005; Naab \& Trujillo 2006). Slowly rotating, round core-S\'ersic
galaxies are preferentially generated through collisionless (gas-poor)
1:1 mergers or generations of re-mergers (e.g., Naab et al.\ 2007;
Burkert et al.\ 2008; Krajnovi\'c et al.\ 2013). The Hopkins
et al.\ (2009) simulations argued that the low ellipticities of
massive galaxies are recovered only if the initial progenitor
spheroids are created from gas-rich (dissipative) mergers 
(see also Naab et al.\ 2013). 

N-body simulations have shown 
that dissipationless major mergers of galaxies involving supermassive
black holes  naturally generate remnants which resemble
core-S\'ersic elliptical galaxies (e.g., Ebisuzaki et al.\ 1991; Milosavljevi\'c \& Merritt 2001; Merritt 2006).
Since it is predominantly  stars on sufficiently radial orbits that pass close to a coalescing black 
hole binary and then get ejected, this causes the deficit of radial orbits and a 
dominance of tangential orbits inside  $R_{b}$ (e.g., Quinlan \& Hernquist 1997;
Milosavljevi\'c \& Merritt 2001; Gebhardt et al.\ 2003; Thomas et
al.\ 2014). Assuming that  this  scattering process ejects stars on  radial
 orbits of all orientation, it suggests that the final isophotes within the galaxy core are likely to be round.
In addition, many authors have pointed out the role of a single central
supermassive black hole in shaping the innermost regions of triaxial
galaxies,
 arguing that the disruption of box orbits passing too close to the black hole
  results in a nearly spherical (round) core (Merritt \& Poon
 2004).  Indeed, as noted in Section~\ref{secV3.2}, the median
 ellipticity of our core-S\'ersic galaxies at
 the break radii ($0.13\pm 0.01$)
 tend to be  more round than the median ellipticity at larger radii
  ($0.20\pm 0.01$), suggestive of i) stellar 
  scattering by a binary supermassive
black hole and/or ii) the
 influence of the central supermassive
black hole after coalescence of the black hole binary.

The sizes and the stellar mass deficits of  depleted galaxy cores 
scale with the mass of the central  supermassive black hole (e.g., Graham
2004; Ferrarese et al.\ 2006; Lauer et al.\ 2007; Rusli et al.\ 2013; Dullo \& Graham
2014). The formation of these cores is predicted to be a cumulative process, such that 
$M_{\rm def} \approx$ $N M_{\rm BH}$ in individual galaxies, where $N$ is the number of
successive major ``dry" mergers (Merritt 2006). In Dullo \& Graham
(2014) we measured central stellar mass deficits typically in the range $0.5 -
4$ $M_{\rm BH}$ for the 24 core-S\'ersic galaxies studied
here. We found that $M_{\rm def}\propto M_{\rm BH}^{3.70\pm 0.76}$ for the sample ensemble. 
Interestingly, as noted in Section~\ref{Sec3.3} (Fig.~\ref{FigV1a}),
galaxies with larger (mass deficit)-to-(black hole mass) ratios possess round cores and round outer 
regions. Given the  predicted relation between this $M_{\rm def}/M_{\rm BH}$ ratio
and the number of major majors (Merritt 2006), this result suggests 
that galaxies which have experienced multiple dissipationless mergers
are round. 

This new observation is in good agreement with the
simulations by Weil \& Hernquist (1996). Further,  Kuehn \&
Ryden (2005, their Figs.\ 6 and 7)  showed that galaxies
in high density environment tend to be round. 
Hao et al.\ (2006) also arrived at
similar conclusions regarding the relation between the ellipticities of
early-type galaxies and their local density.  Likewise, Ryden et
al.\ (2001) studied the relation between galaxy shape and age while
Vincent \& Ryden (2005, see also Alam \& Ryden 2002) examined the
dependence of galaxy shapes on the luminosities.  These studies
concluded that brighter galaxies with old stellar populations, such as
the core-S\'ersic galaxies, are rounder than their fainter
counterparts  which have young stellar populations (see also, e.g., Lauer
et al.\ 2005; Ferrarese et al.\ 2006).  Most recently, Weijmans et al.\ (2014)
 argued that slow rotating early-type galaxies tend to be rounder than fast rotating 
 galaxies (see also Emsellem et al.\ 2007, 2011; Cappellari et al.\ 2011) which contain discs.  
 Here we went one step further and showed for the first time a tendency (albeit only 2 $\sigma$ significance) for galaxies with large
$M_{\rm def}/M_{\rm BH}$ ratios to have round cores ($\epsilon_{\rm inner}
\la 0.2$), while other galaxies with $M_{\rm
  def}/M_{\rm BH} < 2.5$ can have more elliptically-shaped cores ($\epsilon_{\rm
  inner} > 0.2-0.4$) in addition to roundish cores. Using the two-dimensional KS test, we find that  the
   probabilities that  galaxies with $M_{\rm def}/M_{\rm BH} \ge 2.5$  and  $M_{\rm def}/M_{\rm BH} <  2.5$ 
   are drawn from the same parent population is 3.2\%.
   Note that this (weak) trend between the galaxy ellipticity and the $M_{\rm def}/M_{\rm BH}$ ratio is not simply because
    brighter galaxies are rounder (see Section~\ref{Sec3.3}).

In Figs.~\ref{FigV1} (and \ref{FigV5}) we have
illustrated a fair agreement between the break ellipticities
$\epsilon_{\rm inner}$ (and position angles P.A.$_{\rm inner}$) and the
mid-radii ellipticities $<$$\epsilon_{\rm mid}$$>$ (and position angles
$<$P.A.$_{\rm mid}$$>$) for our core-S\'ersic galaxies. This result is
consistent with the notion that core-S\'ersic galaxies are formed
through major ``dry" mergers involving supermassive black holes. That
is, the violent dissipationless relaxation and phase mixing of
progenitor stars (e.g., Naab et al.\ 2006; Burkert et al.\ 2008;
Hopkins et al.\ 2009) together with the chaotic (and possibly
isotropic) stellar scattering by the central supermassive
black hole (e.g., Norman, May,
\& van Albada 1985; Gerhard \& Binney 1985) create a round
core-S\'ersic spheroid with similarly aligned and shaped inner and
outer regions.  Indeed, the bulk (75\%) of our core-S\'ersic galaxies have values for $\epsilon_{\rm inner}$ and 
$<$$\epsilon_{\rm mid}$$>$  which are smaller than $  0.30$ (Fig~\ref{FigV1b}).

In general, Figs.~\ref{FigV1} and \ref{FigV5} agree with
the works by Hao et al.\ (2006) and Chaware et al.\ (2014).  Hao et
al.\ (2006) reported differences in the ellipticity and
position angle between 1 and 1.5 Petrosian half-light radii for nearby
early-type galaxies more elliptical than $\epsilon =0.3$. However,
these differences in the ellipticity and
position angle were statistically less significant for the   
rounder galaxies with $\epsilon < 0.3$ and may have been associated with
the increasing dominance of a disk at large radii. Chaware et al.\ (2014)
 found good agreement between the isophotal
shapes and orientations of the inner (seeing radius to 1.5  Petrosian half-light radius) and intermediate (1.5 Petrosian half-light radius to 3.0 Petrosian half-light radius) regions of 132
distant (0.1 $< z < 0.3$) early-type galaxies, most of which have $-17 < M_{B} < -21$ mag. However,  they reported that these galaxies' shapes
and orientations tend, on average, to change as one goes from
intermediate to larger radii. They concluded that this may be because
of the inclusion of galaxies that are still unrelaxed. Thus, the good agreement between the inner and outer ellipticities and position angles that we find here for our sample of nearby core-S\'ersic galaxies with $M_{V} \la -20.7$ mag (Figs.~\ref{FigV1} and \ref{FigV5}) is generally consistent with the results of  Hao et al.\ (2006) and Chaware et al.\ (2014).

\section{Conclusions}\label{ConV} 

We have used the IRAF {\sc ellipse} task to extract the radial
ellipticity and position angle profiles for a sample of 24 nearby
core-S\'ersic galaxies observed with the {\it HST} WFPC2 and ACS
cameras.  We additionally used the inner ($R\la3\arcsec$) PSF-deconvolved
 ellipticity and position angle profiles  from Lauer et
al.\ (2005)  to
account for the {\it HST} PSF. We found that the PSF did not appear to heavily
affect the ellipticities and position angles at the break radii of our
core-S\'ersic galaxies (Figs.~\ref{FigV2}a, \ref{FigV3} and
\ref{FigV4}). We explored the radial variation of the galaxies' shapes and
orientations.  Our principal conclusions are:

1. For the bulk of the core-S\'ersic galaxies, we measured ellipticities and
position angles at the core-S\'ersic break radii that are in fair agreement with
the average mid-radii ellipticities and position angles determined over
 $R_{\rm e}/2 < R < R_{\rm e}$ (Figs.~\ref{FigV1}, \ref{FigV5}).

2. The median inner ellipticity for our sample is $\epsilon_{\rm med}=
0.13 \pm 0.01$,  rounder than the median ellipticity for the outer regions 
$\epsilon_{\rm med}= 0.20 \pm 0.01$. 13 of our 24 galaxies have inner 
ellipticities ($\epsilon_{\rm inner}$) which are smaller than the mid-radii 
ellipticities  ($<$ $\epsilon_{\rm mid}$$>$), while for 4/24 galaxies $\epsilon_{\rm inner}$ and $<$$\epsilon_{\rm mid}$$>$ are consistent. The remaining 7/24 galaxies have values for $\epsilon_{\rm inner}$ 
 which are greater than $<$$\epsilon_{\rm mid}$$>$ (Figs.~\ref{FigV1}, \ref{FigV1b}). This overall trend may 
indicate the influence of the central supermassive black hole 
in the inner regions of the galaxies. 

3. Core-S\'ersic galaxies with larger (stellar mass deficit)-to-(SMBH
mass), $M_{\rm def}/M_{\rm BH}$ ratios typically have quite round isophotes, suggesting a
possible link between the shapes of the galaxies and their merger
histories. Excluding the only galaxy with an apparent elongated nuclear star cluster, the probability that the ellipticity distributions of galaxies with $M_{\rm def}/M_{\rm BH} \ge 2.5$  and  $M_{\rm def}/M_{\rm BH} <  2.5$ are drawn from the same distribution drops from  3.2\% to 0.3\% (Section~\ref{Sec3.3}).

\section{ACKNOWLEDGMENTS}

This research was supported under the Australian Research Council's
funding scheme (DP110103509 and FT110100263). We are grateful to Paolo
 Bonfini for providing the GALFIT ellipticities and position angles.

\label{lastpage}
\end{document}